\documentclass[prl,aps,floatfix,twocolumn]{revtex4}

\usepackage{graphicx}
\graphicspath{{figures/}}

\usepackage{units,xspace}
\newcommand{\micron}{\ensuremath{\unit{\mu m}}\xspace}

\newcommand{\abs}[1]{\left\vert #1 \right\vert}

\begin{document}

\title{Weak Chaos and Fractional Dynamics in an Optically Driven
  Colloidal Ring}

\author{Yael Roichman}
\affiliation{Department of Physics and Center for Soft Matter
  Research, New York University, New York, NY 10003}

\author{George Zaslavsky}
\affiliation{Department of Physics and Courant Institute, New
  York University, New York, NY 10003}

\author{David G. Grier}
\affiliation{Department of Physics and Center for Soft Matter
  Research, New York University, New York, NY 10003}

\date{\today}

\begin{abstract}
Three colloidal spheres driven around a ring-like optical trap known
as an optical vortex have been predicted to undergo periodic
collective motion due to their hydrodynamic coupling.
In fact, the quenched disorder in the optically-implemented
potential energy landscape
drives a transition to instability evolving into microscopic weak chaos with
fractional dynamics. As a result, the relation between the space-time
selfsimilarity of the system's collective transport properties and its
microscopic weak chaos dynamics is revealed. 
\end{abstract}


\maketitle

Three identical spheres slowly sedimenting through a viscous fluid 
in two or three dimensions generically tumble chaotically
\cite{janosi97}.
When the particles are driven steadily around a ring, by contrast,
their motion is predicted to be purely periodic \cite{reichert04},
with the reduction in dimensionality and the imposition of periodic
boundary conditions \cite{ekieljezewska05} effectively eliminating
the domain of chaotic dynamics.
In this Letter, we demonstrate experimentally 
that chaos can be restored to this system by imposing a small amount
of quenched periodic disorder.
The hydrodynamically coupled spheres' collective motions, furthermore,
are characterized by fractional dynamics.

Our system consists of colloidal silica spheres 1.58~\micron in 
diameter (Bangs Laboratories,Inc. PS04N)  dispersed in a layer of water
40~\micron thick between a glass microscope slide and a coverslip.
Three of these spheres are confined to a horizontal ring 12~\micron
in diameter by a single-beam optical trap known as an optical vortex
\cite{he95,gahagan96,simpson96}.
An optical vortex is formed by bringing a helical
mode of light \cite{allen92} to a focus with a high-numerical-aperture lens.
The helical mode's wavefronts form an $\ell$-fold helix, where
$\ell$ is an integer winding number known as the topological charge.
The axial $\ell$-fold screw dislocation
results in perfect destructive interference along the optical axis, so that
the beam focuses to a ring of light
whose radius is proportional to $\ell$ \cite{curtis03,sundbeck05}.
Polarizable particles are drawn up intensity gradients to the
bright ring, where they are trapped.
Each photon in a helical beam, moreover, carries $\ell \hbar$
orbital angular momentum that can be transferred to trapped
objects \cite{he95a}.
This creates a constant torque that drives the particles around the ring.

\begin{figure}[htbp]
  \centering
  \includegraphics[width=0.8\columnwidth]{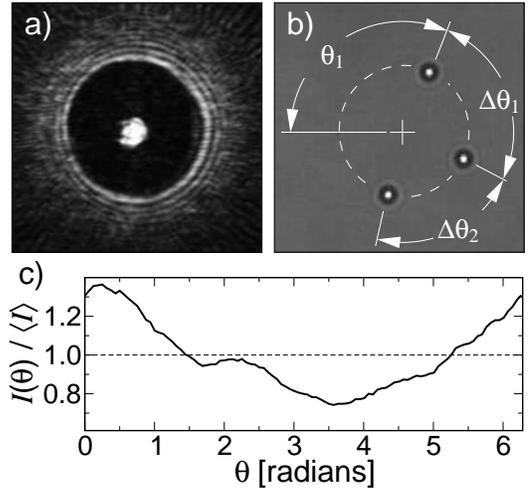}
  \caption{Optically driven colloidal ring.  (a) Projected intensity
pattern for an optical vortex with $\ell = 80$.  (b) Video microscope
image of three colloidal silica spheres trapped on the optical vortex.
(c) Measured intensity variations around the optical vortex's
circumference.}
  \label{fig:vortex}
\end{figure}
Our samples are mounted on the stage of a Nikon TE-2000U inverted
optical microscope, whose objective lens ($100\times$ NA 1.4 oil
immersion Plan-Apo) is used both to project an optical vortex 
and also
to create images of trapped objects.
We imprint helical phase profiles onto the wavefronts of a TEM$_{00}$
beam (Coherent Verdi, $\lambda = 532~\unit{nm}$) using a phase-only
spatial light modulator (Hamamatsu X6750 PPM) in the holographic
optical trapping configuration \cite{dufresne98,curtis02,polin05}.

\begin{figure}[tb]
  \centering
  \includegraphics[width=\columnwidth]{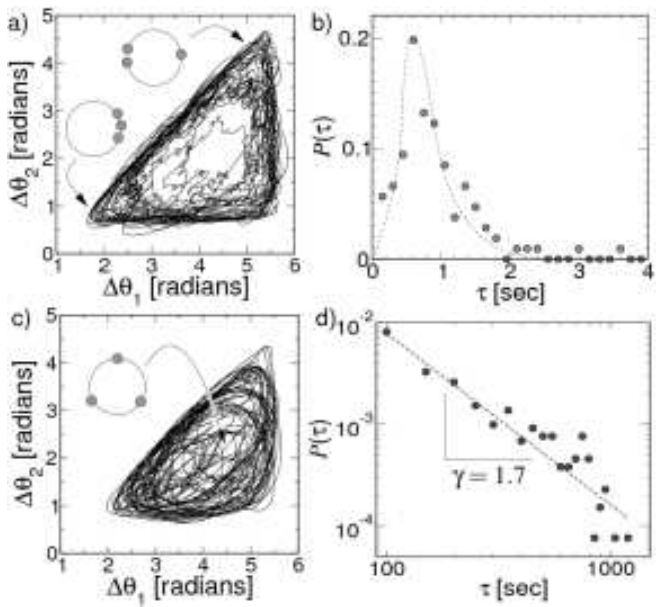}
  \caption{(a) Projection of the phase space reconstructed from
    particle tracks at $\ell = 50$.
    (b) Recurrence time distribution for $\ell = 50$.  The curve is
    a guide to the eye suggesting a cycle period of 0.6~\unit{s}
    (c) Phase space at $\ell = 80$.  (d) Recurrence time distribution
    for $\ell = 80$.}
  \label{fig:phase}
\end{figure}
Figure~\ref{fig:vortex}(a) shows 
an optical vortex with $\ell = 80$ whose image
was obtained by placing a front-surface
mirror in the microscope's focal plane and collecting the reflected
light with the objective lens.
The bright central spot is a conventional optical tweezer formed on
the optical axis by the undiffracted portion of the laser beam.
After adaptively correcting for residual aberrations in the
optical train \cite{polin05,roichman05a}, the
optical vortex is a nearly uniform ring of light with radius
$R_\ell = 12~\micron$.
Figure~\ref{fig:vortex}(b) shows three colloidal spheres trapped on
the ring at 2.5~\unit{W}.  Under these conditions, the particles
circulate once around the ring in $T = 1.2~\unit{s}$.
We track the spheres' angular positions, $\theta_1(t)$, $\theta_2(t)$ and
$\theta_3(t)$, by recording the video stream
at 30 frames per second to a Pioneer 520H-S digital video recorder (DVR) 
with a NEC TI-324AII 
monochrome video camera and extracting their instantaneous positions
at 20~\unit{nm} resolution using standard methods of digital video
microscopy \cite{crocker96}.

The optical vortex's peak intensity, plotted in
Fig.~\ref{fig:vortex}(c) varies by about 20 percent from
the mean around its circumference.
These intensity variations establish an effective potential energy
landscape for the circulating particles resulting from variations in
the local orbital angular momentum flux and from optical gradient
forces \cite{lee06a}.
None of these variations is severe enough to trap a particle,
and speed fluctuations for a single fluctuating particle suggest
that relative mean-square 
variations in the effective potential are smaller than
5 percent \cite{lee06a,ladavac04,pelton04a}.
Structure in the potential energy landscape is
fixed in both position and time and therefore contributes
quenched disorder to the particles' dynamics.

A circulating sphere entrains flows
in the surrounding fluid that
exert forces on the neighboring spheres.
Treating the hydrodynamic coupling in the stokeslet approximation
\cite{pozrikidis92} reveals that the most symmetric state,
with the particles equally spaced around the ring, is linearly
unstable against a dynamical state in which two 
closely spaced spheres outrun the third \cite{reichert04}.
This periodic dynamical state ought to be observed almost
exclusively.
In practice, however, the experimental system can spend
much of its time in the nominally
unstable state.

We characterized the system's dynamics by reconstructing its
phase space from measurements of the independent angular separations, 
$\Delta\theta_1(t) = \theta_2(t) - \theta_1(t)$
and $\Delta\theta_2(t) = \theta_3(t) - \theta_2(t)$, and their associated momenta.
The results in Fig.~\ref{fig:phase}(a) show one continuous trajectory
obtained over 2.5 hours, and smoothed by box-car averaging over ten 
mean circulation periods, $T$.
Smoothing suppresses detailed structure in the phase space
trajectory due to diffusion and
disorder, and thus provides a clearer picture
of the system's intrinsic behavior.

Figure~\ref{fig:phase}(a) shows results from an optical vortex
with $\ell = 50$ and $R_\ell = 8~\micron$
that reveal the predicted periodic orbit \cite{reichert04}.
The parametric trace resembles
a torus, with a pair of spheres racing around
the ring to catch up to the remaining isolated sphere to form
a transient three-particle cluster, whose leading pair
then leaves the remaining sphere behind.  The characteristic
cycle time of $\tau = 0.6~\unit{s}$ substantially exceeds
the 0.3~\unit{s} required for an individual particle to 
circulate once around the ring.

Increasing the topological charge to $\ell = 80$ increases optical
vortex's radius and reduces its mean intensity proportionately.
Both the increased interparticle separation and the decreased 
mean circulation rate weaken the interparticle hydrodynamic coupling.
The quenched disorder is proportional to the local intensity, and thus is
comparatively stronger in the larger optical vortex.
Under these circumstances, the particles' trajectories, plotted in
Fig.~\ref{fig:phase}(c),
make frequent and extended excursions away from the 
nominally stable period cycle into the region of phase space
corresponding to the nominally unstable equilateral configuration.

We characterized this model system's microscopic dynamics by
measuring the distribution of recurrence times, $\tau$
required for a trajectory to revisit regions of phase space
$\delta \theta = 0.5~\unit{rad}$ on a side.
The results for $\ell = 50$ are plotted in Fig.~\ref{fig:phase}(b) and
are consistent with a simple cycle with a period of 0.7~\unit{s}.
Those for $\ell = 80$, plotted in Fig.~\ref{fig:phase}(d), 
display no periodicity, nor are they consistent with
the exponential distribution that would be expected for fully
developed chaos.  
Instead, they follow a power-law decay, with
an exponent $\gamma = 1.7 \pm 0.1$.  

Power-law divergence of trajectories in phase space is a
defining characteristic of weak chaos \cite{zaslavsky05}.
Its emergence in periodically kicked overdamped dynamical systems
has been inferred from measurements on nonlinear Alfven waves in
plasmas \cite{rempel04,burlaga05} and in low-dimensional wall
flow of viscous fluids \cite{jimenez01}.
Unlike these pioneering studies,
we have direct experimental access to the
relevant microscopic degrees of freedom, and therefore do not have
to reconstruct the phase space structure from measurements of
collective properties.

Quenched periodic disorder is known to induce transitions
to chaos in a wide variety of Hamiltonian and non-Hamiltonian systems.
The appearance of weak chaos in a strongly overdamped
system is far less common.
Other overdamped systems displaying noise-induced chaos 
\cite{crutchfield80} rely on thermal forces to 
fully explore phase space.
The optically-driven colloidal ring, by contrast, is 
athermal.

Particles circulating around an
optical vortex are affected by
three factors: the spheres' hydrodynamic interactions, the periodic potential landscape,
and Brownian motion. The importance of each can be inferred by
comparing the measured \cite{crocker96} free-particle
self-diffusion coefficient 
$D_0 = 0.19 \pm 0.02~\unit{\micron^2/s}$ 
with that of a single sphere in an optical vortex \cite{lee06a},
$D_1 = 1.0 \pm 0.2~\unit{\micron^2/s}$, 
and with that of one of the three spheres in the complete system,
$D_3 = 2.0 \pm 0.3~\unit{\micron^2/s}$.
The individual sphere's effective diffusivity exceeds the
free-particle value by a factor of five, demonstrating that
the landscape's quenched disorder has far more influence than
thermal forces.
The fluctuations, moreover, have a non-Gaussian distribution
that reflects the intensity variations around
the optical vortex,
rather than thermal forcing.
This differs from giant amplification of thermal fluctuations
that can occur when particles become marginally trapped
in local potential energy wells \cite{reimann01a,lee06a}.
That fluctuations increase by another factor of two 
with three particles on the
same ring demonstrates that hydrodynamic coupling is 
stronger still.
Because hydrodynamic coupling alone would yield a periodic
orbit \cite{reichert04}, the weakly chaotic dynamics we observe
must be due to the quenched disorder.
Thermal fluctuations do not contribute
appreciably to the microscopic dynamics of the observable
three-particle system, at least in
this range of parameters.

It has long been known that periodic perturbations can induce chaos
in dissipative systems possessing a stable limit cycle
\cite{zaslavsky78}.
Here, hydrodynamic coupling is responsible for the limit cycle
\cite{reichert04} and the combination of viscous damping and
quenched disorder causes the
system to sample other
parts of phase space.
Not any landscape will do this, however.
As suggested in \cite{zaslavsky78} and proved in \cite{wang01,wang02a},
a periodic perturbation can open up a chaotic attractor
near a stable limit cycle if the perturbation possesses 
a fairly broad frequency spectrum,
for instance if it consists of sharp kicks.
The potential energy landscape inferred from Fig.~\ref{fig:vortex}(c)
has this property, kicking the three-particle system three times during each cycle.

\begin{figure}[tb]
  \centering
   \includegraphics[width=\columnwidth]{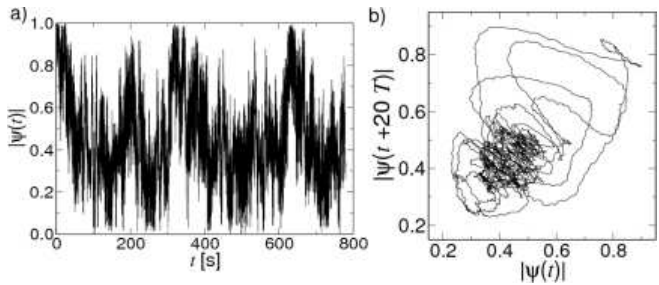}
  \caption{(a) Evolution of the three-fold bond-orientational order
    parameter.  (b) Phase space reconstructed from $\psi(t)$.
  }
  \label{fig:psi}
\end{figure}

Having direct access to a chaotic system's microscopic degrees of
freedom presents a rare opportunity to track
how weak chaos in the microscopic dynamics of an experimental system
influences macroscopic properties.
To show this, we combine the three trajectories into
the complex three-fold bond-orientational order parameter,
\begin{equation}
  \label{eq:psi}
  \psi(t) = \frac{1}{3} \, \sum_{j = 1}^3 \exp\left( 3 i \theta_j(t) \right),
\end{equation}
where $\theta_j$ is measured with respect to a fixed reference direction.
This function's phase tracks the system's rotation about the optical
axis and is useful for measuring
the mean cycle period $T$.
Its magnitude, $\abs{\psi(t)}$, reaches unity when the spheres are evenly spaced and
drops to roughly one third when a pair of spheres is diametrically
opposite from the third.

The trace of $\abs{\psi(t)}$ in Fig.~\ref{fig:psi}(a) is computed for
the trajectory data in Fig.~\ref{fig:phase}(c) and provides a
macroscopic overview of the system's microscopic dynamics, akin to
observing the Brownian motion of a colloidal sphere as a probe of the
microscopic dynamics of the surrounding fluid.
As for other effective macroscopic descriptors, $\psi(t)$ can be used
to reconstruct the underlying microscopic phase space.
For example, Fig.~\ref{fig:phase}(b) shows a Poincare section at
delay $20~T$, which effectively fills the accessible part of the phase space.
The periodic state concentrates the trajectory around $(1/3,1/3)$, while
occasional excursions to the equilateral state fill out the pattern.

\begin{figure}[tb]
  \centering
  \includegraphics[width=0.9\columnwidth]{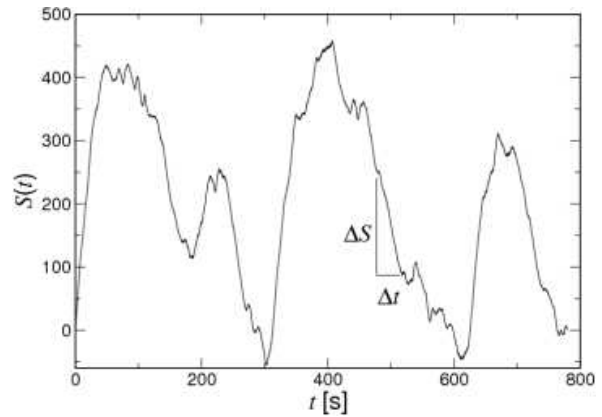}
  \caption{Running sum, $S(t)$, of the three-fold bond-orientational
    order parameter, displaying a hierarchy of jumps $\Delta S$ spanning a
    range of durations $\Delta t$.}
  \label{fig:runningsum}
\end{figure}

\begin{figure}[tb]
  \centering
  \includegraphics[width=\columnwidth]{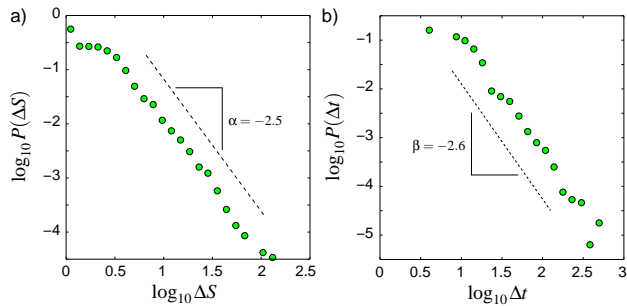}
  \caption{Scaling of the magnitude and duration distributions for jumps in $S(t)$.}
  \label{fig:scaling}
\end{figure}

The order parameter's magnitude also reveals
that the
system switches intermittently between paired and equilateral states.
To analyze the intermittent dynamics we use methods originally
developed to interpret density fluctuation data obtained
from tokamak plasmas \cite{zaslavsky00}.
Given measurements of $\psi(t)$ at $N$ discrete times $t_k$,
the running sum
\begin{equation}
  S(t_n) = \sum_{k=1}^n \abs{\psi (t_k)}
  - \frac{n}{N} \, \left(\abs{\psi (t_N)} - \abs{\psi(t_1)} \right)
\label{sum}
\end{equation}
minimizes the noise in $\abs{\psi(t)}$ and emphasizes
jumps at different scales, as shown in Fig.~\ref{fig:runningsum}.
Jumps in $S(t)$ may be identified as flights
in a system with fractional dynamics.
One then can consider the distribution functions, 
$P_S(\Delta S)$ and $P_t(\Delta t)$, for
the magnitudes, $\Delta S$, and durations $\Delta t$
of these flights.
The corresponding results are presented
in Fig.~\ref{fig:scaling}.

Not only does $\abs{\psi(t)}$ display intermittent jumps, but its
two-state structure appears also to be scale-invariant.
Both the probability distribution for jump magnitudes and jump
durations
are well described by power laws.  The exponents
$\alpha = -2.5 \pm 0.1$ and $\beta = -2.6 \pm 0.2$ for the
magnitude and duration distributions, respectively, are consistent
with each other, and both exceed 2.  This is a signature of
fractional kinetics in the system's chaotic collective dynamics \cite{zaslavsky05}.

Equality of $\alpha$ and $\beta$ arises naturally in any system
undergoing ballistic flights.
For the three colloidal spheres in the present model system, these
flights take the form of random transitions between the equilateral and
periodic dynamical states.
Furthermore, log-periodic oscillations are evident when $S(t)$ is
plotted on a logarithmic time scale.
Such log-periodicity arises for processes characterized by discrete scaling
\cite{zaslavsky05,sornette98} and can be considered as evidence
for hierarchical structure in the system's collective dynamics.
The resulting self-similar structure of weakly chaotic dynamics
is predicted
\cite{zaslavsky05}
to yield a relationship between
the scaling in the distribution of microscopic recurrence times and
the distribution of macroscopic flight durations,
$P(\tau) \, d\tau = P_t(\Delta t) \, d\Delta t$.
This leads to the prediction $\gamma = \alpha - 1$, which is
consistent with our experimental results.

We have demonstrated that imposing a small amount of quenched disorder
on an optically driven colloidal ring
can induce a transition from a periodic steady state to weakly
chaotic dynamics characterized by fractional time-space scaling.
Because the circulating particles in our experiments
repeatedly sample the same fixed optical
intensity pattern, this disorder results in periodic driving with
strong kicks.
Consequently, this provides a model system for studying
disorder-induced transitions to chaos in which both the microscopic
and macroscopic degrees of freedom are experimentally accessible.

This work was supported by the National Science Foundation through
Grant Number DMR-0451589. G.Z. was supported by the ONR Grant
N00014-02-1-0056.  We would like to thank Pablo Jercog for inspiring
discussions and M. Edelman for the help in preparing
Fig.~\ref{fig:scaling}.


\end{document}